\documentclass[aps,prl,twocolumn,amsmath,amssymb,nofootinbib,showpacs,superscriptaddress]{revtex4-1}
\usepackage[english]{babel}
\usepackage{latexsym}
\usepackage{graphics}
\usepackage{graphicx}
\usepackage{epsfig}
\usepackage{color}
\usepackage{bm}
\usepackage{amsmath}
\usepackage{amssymb}
\usepackage{amsthm}
\usepackage{dcolumn}
\usepackage{bm}
\usepackage{float}
\usepackage{hyperref}
\usepackage{color}
\usepackage{epstopdf}
\usepackage{cleveref}
\usepackage[svgnames]{xcolor}
\hypersetup{hidelinks,colorlinks=true,allcolors=DarkBlue}

\begin{document}

\title{Novel $p$-wave superfluids of fermionic polar molecules}

\author{A.K. Fedorov}
\affiliation{Russian Quantum Center, Skolkovo, Moscow Region 143025, Russia}
\affiliation{LPTMS, CNRS, Univ. Paris-Sud, Universit\'e Paris-Saclay, Orsay 91405, France}

\author{S.I. Matveenko}
\affiliation{L.D. Landau Institute for Theoretical Physics, Russian Academy of Sciences, Moscow 119334, Russia}
\affiliation{LPTMS, CNRS, Univ. Paris-Sud, Universit\'e Paris-Saclay, Orsay 91405, France}

\author{V.I. Yudson}
\affiliation{Russian Quantum Center, Skolkovo, Moscow Region 143025, Russia}
\affiliation{Institute for Spectroscopy, Russian Academy of Sciences, Troitsk, Moscow 142190, Russia}

\author{G.V. Shlyapnikov}
\affiliation{Russian Quantum Center, Skolkovo, Moscow Region 143025, Russia}
\affiliation{LPTMS, CNRS, Univ. Paris-Sud, Universit\'e Paris-Saclay, Orsay 91405, France}
\affiliation{\mbox{Van der Waals-Zeeman Institute, University of Amsterdam, Science Park 904, 1098 XH Amsterdam, The Netherlands}}
\affiliation{Wuhan Institute of Physics and Mathematics, Chinese Academy of Sciences, Wuhan 430071, China}

\date{\today}
\begin{abstract}
Recently suggested subwavelength lattices offer remarkable prospects for the observation of novel superfluids of fermionic polar molecules. 
It becomes realistic to obtain a topological $p$-wave superfluid of microwave-dressed polar molecules in 2D lattices at temperatures of the order of tens of nanokelvins, 
which is promising for topologically protected quantum information processing.
Another foreseen novel phase is an interlayer $p$-wave superfluid of polar molecules in a bilayer geometry.

\end{abstract}
\maketitle

\subsection{Introduction}

Non-conventional superconductors and superfluids attract a great deal of interest due to their non-trivial transport properties and/or topological behavior 
\cite{Volovik,Mineev,Nelson,Rice,Read,DasSarma,Bulaevskii,Graf,FFLO,FFLO2,FFLO3}.
This behavior has been actively discussed in two dimensions (2D) for the $p_x+ip_y$ superfluid of identical fermions, 
where Cooper pairs have orbital angular momentum equal to unity~\cite{Gurarie,Nishida,Gurarie2,Gurarie3,Shlyapnikov,Shlyapnikov2}.
Quantized vortices in this superfluid carry zero-energy Majorana modes on their cores~\cite{Read,Stern,Gurarie4}.
These modes cause the vortices to obey non-Abelian exchange statistics, 
which is a basis for topologically protected quantum information processing~\cite{Nayak,Nayak2}. 
However, 
the $p$-wave topological superfluid of ultracold atoms is either collisionally unstable near a Feshbach resonance, 
or has a vanishingly low superfluid transition temperature far from the resonance~\cite{3body,3body2,3body3}.

Successful experiments on the creation of ground-state ultracold polar molecules \cite{Ye,Gabbanini,Ye2,Ye3,Ye4,Ye5,Ye6,Ye7,Jin,Weidemuller,Ketterle2,Zwierlein} 
opened fascinating prospects for obtaining non-conventional superfluids \cite{DipoleRevs,DipoleRevs2,DipoleRevs3}. 
In particular, microwave-dressed polar molecules confined to 2D may acquire an attractive dipole-dipole tail in the interaction potential, 
which ensures the emergence of collisionally stable $p$-wave superfluid with a reachable transition temperature \cite{Shlyapnikov,Shlyapnikov2}. 
Another interesting system concerns fermionic polar molecules in a bilayer geometry. 
Here they may form interlayer superfluids in which Cooper pairs consist of molecules belonging to different layers \cite{Santos,Baranov,Santos2,Zinner}. 

\begin{figure}[t] 
\begin{centering}
\includegraphics[width=0.477\textwidth]{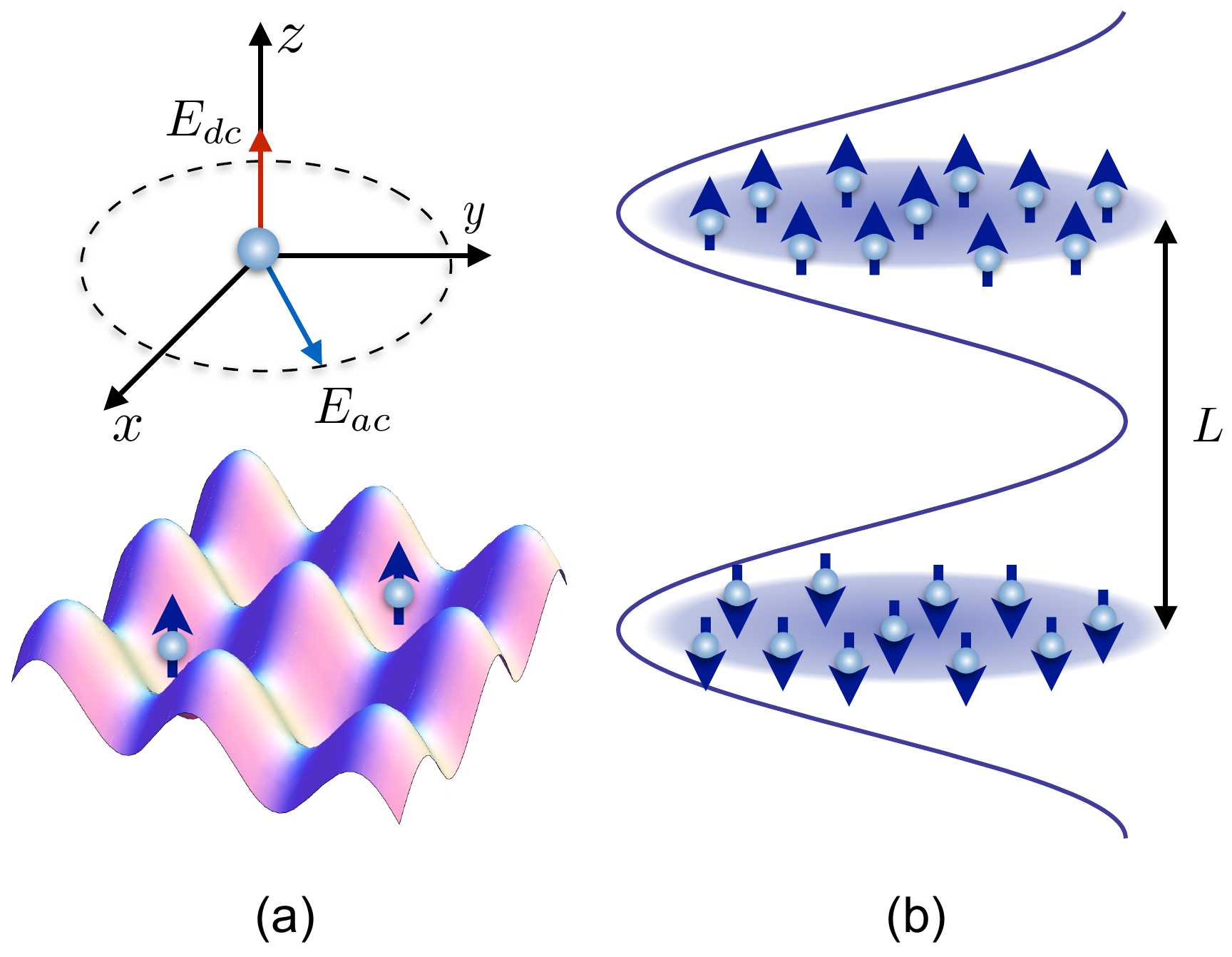}
\end{centering}
\vskip -4mm
\caption
{
Setups for $p$-wave superfluids of polar molecules:
(a) polar molecule in an external microwave field $E_{ac}$ rotating in the plane perpendicular to the stationary field $E_{dc}$ (upper part), and microwave-dressed polar molecules 
loaded in a 2D lattice (lower part);
(b) bilayer system of polar molecules with dipole moments in the upper and lower layers, oriented opposite to each other.
}
\label{fig:main}
\end{figure}

It this paper we consider novel $p$-wave superfluids of fermionic polar molecules in 2D lattice geometries (see Fig.~\ref{fig:main}).
It is shown that a collisionally stable topological $p_x+ip_y$ 
superfluid of identical microwave-dressed polar molecules may emerge in a 2D lattice due to a long-range character of the dipole-dipole interaction.
We also show how one can get a $p$-wave interlayer superfluid of fermionic polar molecules in a bilayer geometry,
which can be a quantum simulator of superconductivity in layered condensed matter systems \cite{Bulaevskii,Graf}. 
 It is crucial to rely on the recently proposed subwavelength lattices \cite{Zoller4,Berman,Berman2,Dalibard,Lukin3,Cirac,Chen,Spreeuw,Lukin2,Kimble,Kimble2}, 
where the lattice constant (interlayer spacing in the bilayer system) can be as small as about 50 nm. 
An increase of energy scales in such lattices makes it realistic to obtain sizeable transition temperatures of the order of tens of nanokelvins.  

\subsection{General relations and qualitative arguments}

The superfluid pairing of identical fermions is characterized by the order parameter $\Delta({\bf r},{\bf r}\,')=V({\bf r}-{\bf r}\,')$ $\times\langle{\hat\psi({{\bf r}}\,')\hat\psi({{\bf r}})}\rangle$,
where $V({\bf r}-{\bf r}\,')$ is the interaction potential,
the symbol $\langle...\rangle$ denotes the statistical average, 
and $\hat\psi({{\bf r}})$ is the field operator of fermions.
For spin-$1/2$ fermions one of the field operators in the expression for $\Delta({\bf r},{\bf r}\,')$ is for spin-$\uparrow$ fermions,
and the other one for spin-$\downarrow$ fermions.
In free space the order parameter depends on the coordinates ${\bf r}$ and ${\bf r}\,'$ only through the difference $({\bf r}-{\bf r}\,')$.
In 2D the transition temperature $T_c$ of a Fermi gas from the normal to superfluid regime is set by the Kosterlitz-Thouless transition. 
However, for a weak attractive interaction the order parameter and the superfluid transition temperature can be found in the BCS approach \cite{Miyake}.
For both spinless and spin-$1/2$ fermions the renormalized gap equation for the order parameter in the momentum space, 
$\Delta_{\bf{k}}{=}\int{d^2{r}\Delta({\bf r}-{\bf r}\,')\exp\left[{i\bf{k}\left({\bf r}-{\bf r}\,'\right)}\right]}$,
reads (see \cite{Shlyapnikov,Shlyapnikov2} and references therein):
\begin{equation}\label{eq:gap}
	\begin{split}
	\Delta_{\bf k}{=}&{-}{\int}{\frac{d^2k'}{(2\pi)^2}f({\bf k}',{\bf k}){\Delta_{{\bf k'}}\!}{\left\{{\!\mathcal{K}(k'){-}\frac{1}{2(E_{k'}{-}E_k{-}i0)}\!}\right\}}} \\
	&-\int\frac{d^2k'}{(2\pi)^2}\delta{V}({\bf k}',{\bf k})\Delta_{\bf k'}\mathcal{K}(k'),
	\end{split}
\end{equation}
where $f({\bf k}',{\bf k})$ is the off-shell scattering amplitude and $E_k=\hbar^2k^2/2m$ with $m$ being the particle mass.
The single particle excitation energy is given by $\epsilon_k=$$\sqrt{(E_k-\mu)^2+|\Delta(k)|^2}$
where $\mu$ is the chemical potential, and $\mathcal{K}(k)=\tanh(\epsilon_k/2T)/2\epsilon_k$.
For weak interactions chemical potential coincides with the Fermi energy $E_F=\hbar^2k_F^2/2m$ ($k_F$ is the Fermi momentum). 
The quantity $\delta V({\bf k}',{\bf k})$ is a correction to the bare interparticle interaction due to polarization of the medium by colliding particles. 
The leading terms of this quantity introduced by Gor'kov and Melik-Barkhudarov \cite{GMB}, are second order in the bare interaction (see Methods).  

In order to gain insight in what is happening, we first omit the correction $\delta V({\bf k}',{\bf k})$ in Eq.~(\ref{eq:gap}).
We then put $k=k_F$, and notice that the main contribution to the integral over $k'$ in Eq.~(\ref{eq:gap}) comes from $k'$ close to $k_F$. 
At temperatures $T$ tending to the critical temperature $T_c$ from below, we put $\epsilon_{{\bf k}'}{=}|E_{k'}-E_F|$ in $\mathcal{K}(k')$. 
For the pairing channel related to the interaction with orbital angular momentum $l$, this immediately leads to an estimate:  
\begin{equation}\label{eq:Tc}
	T_c\sim E_F\exp\left[-\frac{1}{\rho(k_F)|f_l(k_F)|}\right]. 
\end{equation}
The quantity $\rho(k_F)=m/2\pi\hbar^2$ in the exponent of Eq. (\ref{eq:Tc}) is the density of states on the Fermi surface, 
and $f_l(k_F)$ is the on-shell scattering amplitude.  

In the lattice with a period $b$ satisfying the condition $k_Fb\ll1$, 
the superfluid paring of fermions can be considered as that of particles with effective mass $m^{*}>m$ in free space.
The density of states $\rho(k_F)$ is then given by the same expression, with $m$ replaced by $m^{*}$.
Thus, the BCS exponent $[\rho(k_F)|f_l(k_F)|]^{-1}$ in the lattice is smaller than in free space at the same $k_F$ (density) if there is no significant reduction in the scattering amplitude. 
Hence, although the Fermi energy decreases by the same factor $m/m^*$, 
the critical temperature $T_c$ in the lattice can be much larger than in free space.    
This is the case for the $s$-wave pairing of short-range interacting spin-$1/2$ fermions in the tight binding model,
if the extension of the particle wavefunction in the lattice site greatly exceeds the characteristic radius of the interparticle interaction. 
An increase of the critical temperature for the $s$-wave superfluidity by the lattice potential has been indicated in Refs.~\cite{Lukin,Ketterle}. 

The situation changes for the $p$-wave pairing of identical fermions attractively interacting via a short-range potential. This pairing in an optical lattice at very low temperatures has been considered in Ref. \cite{Iskin} (more sophisticated lattice models, where $p$-wave pairing is constructed with the use of $s$-wave pairing at intermediate stages, were recently suggested in Refs.~\cite{Buchler,Liu}).
In the tight binding model two such fermions can not be in the same lattice site unless one of them occupies a higher Bloch band. 
Therefore, the main contribution to the scattering amplitude comes from the interaction between two fermions sitting in neighboring sites \cite{Iskin}. 
In particular, 
the fermions undergo quantum tunneling from the centers of their sites and experience the short-range interaction in the spatial region where their wavefunctions are attenuated. 
This strongly suppresses the interaction amplitude and leads to a very low critical temperature.
We however show below that the picture is drastically different for an attractive long-range interaction between the fermions.

\subsection{$P$-wave pairing of microwave-dressed \\ polar molecules in a 2D lattice}

We will consider identical fermionic polar molecules in a 2D lattice of period $b$. 
Being dressed with a microwave field, they acquire an attractive dipole-dipole tail in the interaction potential \cite{Zoller,Gorshkov3,Shlyapnikov,Shlyapnikov2}:  
\begin{equation}\label{eq:V}
	V(r)=-d^2/r^3.
\end{equation}
Here $d$ is an effective dipole moment, and we assume that Eq.~(\ref{eq:V}) is valid at intermolecular distances $r\gtrsim{b}$. 
This leads to superfluid $p$-wave pairing of the molecules. 
In free space the emerging ground state is the topological $p_x+ip_y$ superfluid, and the leading part of the scattering amplitude can be obtained in the first Born approximation \cite{Shlyapnikov,Shlyapnikov2}.   
We assume the weakly interacting regime at a small filling factor in the lattice, $k_Fb\ll{1}$.

The Hamiltonian of the system is $\hat{\mathcal{H}}=\hat{H}_{0}+\hat{H}_{\mathrm{int}}$, with
\begin{equation}\label{eq:H0lat}
	\hat{H}_{0}=\sum\nolimits_{\bf q}{\varepsilon_{\bf q}\hat{a}_{\bf q}^{\dagger}\hat{a}_{\bf q}^{}},
\end{equation}
where $\hat{a}_{\bf q}^{}$, $\hat{a}_{\bf q}^{\dagger}$ are the annihilation and creation operators of a molecule with quasimomentum ${\bf q}$, 
and $\varepsilon_{\bf q}$ is the single particle energy. 
In the low momentum limit we have $\varepsilon_{\bf q}=\hbar^2q^2/2m^{*}$,
where $m^{*}>m$ is the effective mass in the lowest Bloch band. 
The quantity $\hat{H}_{\mathrm{int}}$ describes the interaction between the molecules and is given by
\begin{equation}\label{eq:Hitnlat}
	\hat{H}_{\mathrm{int}}=
	-\frac{1}{2}\sum_{{\bf r}_j\ne{\bf r}'_j}{\hat\psi^{\dagger}({\bf r}_j)\hat\psi^{\dagger}({\bf r}'_j)\frac{d^2}{|{\bf r}_j-{\bf r}'_j|^3}\hat\psi^{}({\bf r}'_j)\hat\psi^{}({\bf r}_j)},
\end{equation}
where $\hat\psi^{}({\bf r}_j)$ is the field operator of a particle in the lattice site $j$ located at ${\bf r}_j$ in the coordinate space. 
At a small filling factor in the low momentum limit, 
the main contribution to the matrix elements of $\hat{H}_{\mathrm{int}}$ comes from intermolecular distances $|{\bf r}_j-{\bf r}'_j|\gg{b}$ (see Methods). 
Therefore, we may replace the summation over ${\bf r}_j$ and ${\bf r}'_j$ by the integration over $d^2{\bf r}_j$ and $d^2{\bf r}'_j$.
As a result the Hamiltonian of the system reduces to 
\begin{equation}\label{eq:H}
	\begin{split}
	\hat{H}=&-\int{\frac{\hbar^2}{2m^{*}}\hat\psi^{\dagger}({\bf r})\nabla^2\hat\psi^{}({\bf r})d^2r}\\
	&-\frac{d^2}{2}\int{\frac{1}{|{\bf r}-{\bf r}'|^3}\hat\psi^{\dagger}({\bf r})\hat\psi^{\dagger}({\bf r}')\hat\psi^{}({\bf r}')\hat\psi^{}({\bf r})d^2r},
	\end{split}
\end{equation}
where the first term in the right hand side is $\hat{H}_{0}$ (\ref{eq:H0lat}) rewritten in the coordinate space.
We thus see that the problem becomes equivalent to that of particles with mass $m^{*}$ in free space.

The scattering amplitude at $k=k_F$, 
which enters the exponential factor in Eq.~(\ref{eq:Tc}), is obtained from the solution of the scattering problem in the lattice. 
For particles that have mass $m^*$ (see Methods),
the amplitude is written as follows
\begin{equation}\label{eq:scatt}
	f(k_F){=}-\frac{8}{3}\frac{\hbar^2}{m^*}k_Fr^{*}_{\rm eff}+\frac{\pi}{2}\frac{\hbar^2}{m^*}(k_Fr^{*}_{\rm eff})^2\ln\left({Bk_Fr^{*}_{\rm eff}}\right),
\end{equation}
where $k_Fr^{*}_{\rm eff}\ll1$, and $B$ is a numerical coefficient coming from short-range physics.  
Since for weak interactions two fermions practically do not get to the same lattice site, 
for calculating $B$ we may introduce a perfectly reflecting wall at intermolecular distances $r\sim b$ (see Methods). 
For the superfluid pairing the most important are particle momenta ${\sim}k_F$. 
Therefore, the low-momentum limit requires the inequality $k_Fb\ll 1$.

The solution of the gap equation (\ref{eq:gap}) then leads to the $p_x+ip_y$ superfluid with the critical temperature (see Methods):
\begin{equation}\label{eq:TcP}
	T_c=E_F\frac{\kappa}{(k_Fr^{*}_{\rm eff})^{9\pi^2/64}}\exp\left[-\frac{3\pi}{4k_Fr^{*}_{\rm eff}}\right],
\end{equation}
where the coefficient $\kappa$ is related to $B$ and depends on the ratio $r^*_{\rm eff}/b$ (see Methods). 
There are two important differences of equation (\ref{eq:TcP}) from a similar equation in free space obtained in Ref. \cite{Shlyapnikov}. 
First, the Fermi energy $E_F$ is smaller by a factor of $m/m^*$, and  
the effective dipole-dipole distance $r^*_{\rm eff}$ is larger than the dipole-dipole distance in free space by $m^*/m$. 
Second, the coefficient $B$ and, hence, 
$\kappa$ in free space is obtained from the solution of the Schr\"odinger equation in the full microwave-induced potential of interaction between two molecules, 
whereas here $B$ follows from the fact that the relative wavefunction is zero for $r\leq b$ (perfectly reflecting wall). 

It is clear that for the same 2D density $n$ (and $k_F$) the critical temperature in the lattice is larger than in free space because the BCS exponent in Eq.~(\ref{eq:TcP}) is smaller.
However, in ordinary optical lattices one has the lattice constant $b\gtrsim 200$ nm.
In this case, for $m^*/m\approx 2$ 
(still the tight binding case with $b/\xi_0\approx 3$, where $\xi_0$ is the extension of the particle wavefunction in the lattice site) 
and at a fairly small filling factor (let say, $k_Fb=0.35$)
the Fermi energy for the lightest alkaline polar molecules NaLi is about $10$ nK ($n\approx{2\times10^{7}}$ cm$^{-2}$). 
Then, even for $k_Fr^*_{\rm eff}$ approaching unity the critical temperature is only of the order of a nanokelvin 
(for $k_Fb=0.35$ and $r^*_{\rm eff}/b\approx{3}$ Fig. \ref{fig2} in Methods gives $\kappa\sim1$).

The picture is quite different in recently introduced subwavelength lattices \cite{Zoller4,Berman,Berman2,Dalibard,Lukin3,Cirac,Chen,Spreeuw,Kimble},
where the lattice constant can be as small as $b\simeq{50}$ nm. 
This strongly increases all energy scales, and even for a small filling factor the Fermi energy may become of the order of hundreds of nanokelvins. 
Subwavelength lattices can be designed using 
adiabatic dressing of state-dependent lattices \cite{Zoller4}, 
multi-photon optical transitions \cite{Berman,Berman2},
spin-dependent optical lattices with time-dependent modulations \cite{Dalibard},
as well as nanoplasmonic systems \cite{Lukin3},
vortex arrays in superconducting films \cite{Cirac},
periodically patterned graphene monolayers \cite{Chen},
magnetic-film atom chips \cite{Spreeuw},
and photonic crystals \cite{Lukin2,Kimble,Kimble2}. 
These interesting proposals already stimulated studies related to many-body physics in such lattices, 
in particular the analysis of the Hubbard model and engineering of spin-spin Hamiltonians \cite{Kimble}. 

In the considered case of $p_x+ip_y$ pairing in the 2D lattice, 
putting $b=50$ nm, for the same $k_Fb$ as above the Fermi energy for NaLi molecules exceeds 200 nK ($n\approx{4\times10^{8}}$ cm$^{-2}$).
Then, for the same $\kappa\sim1$ and $k_Fr^*_{\rm eff}$ approaching unity we have $T_c\sim{20}$ nK, which is twice as high as in free space. 
An additional advantage of the lattice system is the foreseen quantum information processing, since addressing qubits in the lattice is much easier than in free space.  

Note that there is a (second-order) process, 
in which the interaction between two identical fermions belonging to the lowest Bloch band provides a virtual transfer of one of them to a higher band.
Then, the two fermions may get to the same lattice site and undergo the inelastic process of collisional relaxation. 
The rate constant of this second-order process is roughly equal to the rate constant in free space, 
multiplied by the ratio of the scattering amplitude (divided by the elementary cell area) to the frequency of the potential well in a given lattice site 
(the difference in the energies of the Bloch bands). 
This ratio originates from the virtual transfer of one of the fermions to a higher band and does not exceed $(\xi/b)^2$. 
Even in not a deep lattice, where $m^{*}/m$ is 2 or 3, we have $(\xi/b)^2<0.1$. 
Typical values of the rate constant of inelastic relaxation in free space are $\sim{10^{-8}{-}10^{-9}}$ cm$^2$/s ~\cite{Shlyapnikov}, 
and hence in the lattice it will be lower than $10^{-9}$ or even $10^{-10}$~cm$^2$/s. 
Thus, the rate of this process is rather low and for densities approaching $10^{9}$~cm$^{-2}$ the decay time will be on the level of seconds or even tens of seconds.

\subsection{Interlayer $p$-wave superfluid \\ of fermionic polar molecules in a bilayer system}

Another interesting novel superfluid of fermionic polar molecules is expected in a bilayer system, 
where dipoles are oriented perpendicularly to the layers and in opposite directions in different layers. 

Such a bilayer configuration, but with all dipoles oriented in the same direction, has been considered in Refs. \cite{Santos,Baranov,Santos2,Zinner}. 
As found, it should form an interlayer $s$-wave superfluid, where Cooper pairs are formed by dipoles of different layers due to the $s$-wave dipolar interaction between them. 

For the dipoles of one layer that are opposite to the dipoles of the other one, the picture of interlayer pairing is different. 
The $s$-wave pairing is practically impossible, and the system may form $p$-wave and higher partial wave superfluids. 
This type of bilayer systems can be created by putting polar molecules with rotational moment $J=0$ in one layer, and molecules with $J=1$ in the other.
Then, applying an electric field (perpendicular to the layers) one gets a field-induced average dipole moment of $J=0$ molecules parallel to the field, 
and the dipole moment of $J=1$ molecules oriented in the opposite direction. 
One should also prevent a flip-flop process in which the dipole-dipole interaction between given $J=1$ and $J=0$ molecules reverses their dipoles, 
thus inducing a rapid three-body decay in collisions of a dipole-reversed molecule with two original ones. 
This can be done by making the electric field inhomogeneous, 
so that it is larger in the layer with $J=0$ molecules and the flip-flop process requires an increase in the Stark energy.
This process will be suppressed if the difference in the Stark energies of molecules in the layers significantly exceeds the Fermi energy,
which is a typical kinetic energy of the molecules ($\sim100$ nK for the example considered below).
This is realistic for present facilities.

For the dipole moment close to 1 Debye and the interlayer spacing of 50 nm, 
one thus should have the field gradient (perpendicularly to the layers) significantly exceeding 0.5 kV/cm$^2$. 
This could be done by using electrodes consisting of four rods, and even a higher gradient $\sim 30$ kV/cm$^2$ should be achievable \cite{Bohn,Hoekstra}. 
By changing the positions of the rods one can obtain the field gradient exceeding 0.5 kV/cm$^2$ in the direction perpendicular to the layers of the bilayer system.
The field itself will not be exactly perpendicular to the layers and there will also be the field gradient parallel to the layers. 
This, however, does not essentially influence the physics. 

The potential of interaction between two molecules belonging to different layers has the form: 
\begin{equation}\label{eq:potential}
	V_L(r)=-d^2{(r^2-2L^2)}/{(r^2+L^2)^{5/2}},
\end{equation}
where $L$ is the interlayer spacing, $r$ is the in-layer separation between the molecules, and $-d^2$ is the scalar product of the average dipole moments of these molecules. 
The potential $V_L(r)$ is repulsive for $r<\sqrt{2}L$ and attractive at larger $r$. 
The potential well is much more shallow than in the case of all dipoles oriented in the same direction, which was considered in Refs. \cite{Santos,Baranov,Santos2}. 
We have checked that $s$-wave interlayer dimers, which exist at any $r_*/L$, are weakly bound even for $r_*/L\approx 3$. 
Their binding energy at $r_*/L\lesssim 3$ is much smaller than the Fermi energy at least for $k_FL>0.1$. 
For such $r_*/L$, interlayer dimers with orbital angular momenta $|l|\geq 1$ do not exist. 
We thus are dealing with purely fermionic physics.

For the analysis of the superfluid pairing we are interested in particle momenta $k\sim k_F$. 
As well as in the case of all dipoles oriented in the same direction  \cite{Santos,Baranov,Santos2,Zinner}, 
under the condition $k_Fr^*{\ll}1$ (where $r^*{=}md^2/\hbar^2$) the amplitude of interlayer interaction is obtained in the Born approximation. 
The Fourier transform of the potential~(\ref{eq:potential}) is
\begin{equation}\label{eq:potential2}
	V_L({\bf k}',{\bf k})=(2\pi\hbar^2/m)r^{*}|{\bf k}-{\bf k}'|\exp\left[{-{|{\bf k}-{\bf k}'|}L}\right],
\end{equation}
and in the first Born approximation the on-shell amplitude of the $l$-wave scattering at $k=k_F$ reads (see Methods): 
\begin{equation}\label{eq:scattering}
\begin{split}
	f_l(k_F)&=\frac{2\hbar^2k_Fr^{*}}{m}{\int_0^{2\pi}}d\phi\cos(l\phi)\\
	&\times |\sin\left({\phi}/{2}\right)|\exp\left[{-2{k_F}L|\sin\left({\phi}/{2}\right)|}\right].
\end{split}
\end{equation}
The $s$-wave amplitude is positive, i.e. the $s$-wave channel corresponds to repulsion. 
Note that for extremely low collision energies comparable with the dimer binding energy, where the Born approximation is not accurate, the $s$-wave scattering amplitude can be negative. 
This, however, does not lead to superfluid $s$-wave pairing.

The channels with $|l|\geq{1}$ correspond to attraction. 
A straightforward calculation shows that for $k_FL\lesssim0.7$ the largest is the $p$-wave amplitude and, hence, 
at sufficiently low temperatures the system will be an interlayer $p$-wave superfluid. 
As for $d$-wave and higher partial wave superfluids, they are possible only at extremely low temperatures. 
Thus, we confine ourselves to the $p$-wave pairing and employ the BCS approach. 

A detailed analysis of the gap equation (\ref{eq:gap}), 
which includes first and second order contributions to the scattering amplitude and Gor'kov-Melik-Barkhudarov corrections, is given in Methods. 
The critical temperature for the $p$-wave superfluidity proves to be (see Methods):
\begin{equation}\label{Tcp}
	T_{c}=E_F\beta(k_FL)\exp\left[-\frac{F(k_FL)}{k_Fr^*}\right],
\end{equation}
and for not very small $k_Fr^{*}$ the validity of the perturbative treatment of the Gor'kov-Melik-Barkhudarov corrections requires $k_FL\gtrsim{0.15}$ (see Methods).
The functions $F(k_FL)$ and $\beta(k_FL)$ are given in Methods. For $k_FL$ ranging from 0.15 to 0.3 the function $F$ increases from 3.4 to 5, 
and the coefficient $\beta$ is fairly large, being about $80$ at $k_FL=0.15$ (see Fig. \ref{fig3} and Methods).

Creating the bilayer system by using a 1D subwavelength lattice we may have $L\approx50$ nm.
In this case, for $k_FL=0.15$ the Fermi energy of NaLi molecules is close to $100$ nK, and the critical temperature for ${k_Fr^{*}}$ approaching $0.5$ is about~$10$ nK.

For completeness, we also consider the regime of strong interactions within a single layer.
Assuming that the coupling between the layers is still fairly weak, 
we have superfluid (interlayer) pairing between {\it quasiparticles}. 
Related problems have been discussed for coupled 2D Fermi liquids as models for layered superconductors \cite{Graf}.
In this case, we replace the bare mass $m$ by the effective mass $m^{*}$ and account for renormalization of the fermionic Green functions by a factor $Z<1$ \cite{AGD}.
Then, the expression for the transition temperature takes the form:
\begin{equation}\label{TcSI}
	T_{c}\sim{E_F}\exp\left[-\frac{F(k_FL)}{k_Fr^*}\frac{m}{m^{*}}\frac{1}{Z^2}\right],
\end{equation}
where we can not determine the pre-exponential coefficient. 
Therefore, Eq. (\ref{TcSI}) only gives an order of magnitude of $T_c$. 
For $k_FL=0.3$ and $L\approx 50$ nm the Fermi energy of NaLi molecules is about 400 nK, and for, let say, $k_Fr_*\approx 2$ the dimer physics is still not important. 
Then, using the effective mass and factor $Z$ from the Monte Carlo calculations \cite{Giorgini} one may think of superfluid transition temperatures of the order of several tens of nanokelvins.

\subsection{Conclusions}

We have demonstrated the emergence of the topological $p_x+ip_y$ superfluid for identical microwave-dressed fermionic polar molecules in a 2D lattice.
Another novel $p$-wave superfluid is found to emerge for fermionic molecules in a bilayer system, 
with dipoles of one layer opposite to the dipoles of the other one.
In both cases the use of subwavelength lattices with a period $b\simeq{50}$ nm 
(creation of the bilayer system with the interlayer spacing $L\simeq{50}$ nm) 
allows one to obtain superfluid transition temperature of the order of tens of nanokelvins.
This opens interesting prospects for topologically protected quantum information processing with $p_x+ip_y$ superfluids in 2D lattices.
The interlayer $p$-wave superfluid in bilayer systems, together with the earlier proposed $s$-wave interlayer superfluid \cite{Santos,Baranov,Santos2,Zinner} 
and superfluids in multilayer fermionic systems \cite{Potter},
can be a starting point for the creation of more sophisticated layered structures. 

Superfluidity itself can be detected in the same way as in the case of $s$-wave superfluids \cite{GPS, Zwierlein3}. 
Rotating the $p_x+ip_y$ superfluid and inducing the appearance of vortices 
one can find signatures of Majorana modes on the vortex cores in the RF absorption spectrum \cite{Cooper2}.
Eventually, one can think of revealing the structure of the order parameter by visualizing vortex-related dips in the density profile
on the approach to the strongly interacting regime,
where these dips should be pronounced at least in time-of-flight experiments. 

\subsection{Methods}

\subsubsection{Scattering problem and superfluid pairing of \\ microwave-dressed polar molecules in a 2D lattice}

As we concluded in the main text, in the low momentum limit at a small filling factor the system of lattice polar molecules is 
equivalent to that of molecules with effective mass $m^{*}$ in free space.
We now demonstrate this explicitly by the calculation of the off-shell scattering amplitude $f({\bf k}',{\bf k})$.
For our problem the main part of the scattering amplitude can be obtained in the Born approximation \cite{Shlyapnikov}.  

\begin{figure*}[t]
\includegraphics[width=2\columnwidth]{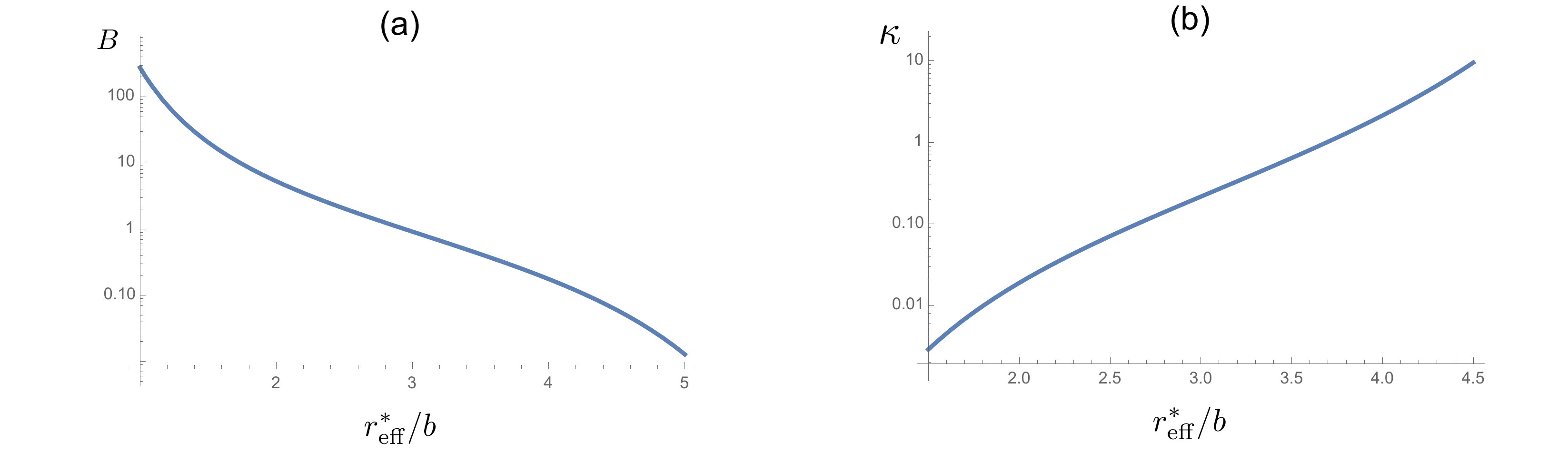}
\vskip -5mm
\caption
{
	Coefficients $B$ and $\kappa$ as functions of $r^{*}_{\rm eff}/b$.
}
\label{fig2}
\end{figure*}

In the lattice the scattering amplitude is, strictly speaking, the function of both incoming quasimomenta ${\bf q}_1,{\bf q}_2$ and outgoing quasimomenta ${\bf q}'_1,{\bf q}'_2$. 
However, in the low-momentum limit where ${qb}\ll{1}$, 
taking into account the momentum conservation law the amplitude becomes the function of only relative momenta ${\bf k}=({\bf q}_1-{\bf q}_2)/2$ and ${\bf k}'=({\bf q}'_1-{\bf q}'_2)/2$. 
For the off-shell scattering amplitude the first Born approximation gives:
\begin{equation}\label{Bf} 
\begin{split}
	f&({\bf k}',{\bf k}){=}S{\int}\chi^*_{{\bf q}'_1}({\bf r}_1)\chi^*_{{\bf q}'_2}({\bf r}_2)V({\bf r}_1-{\bf r}_2)\\
	&\times\chi_{{\bf q}_1}({\bf r}_1)\chi_{{\bf q}_2}({\bf r}_2)d^2r_1d^2r_2 \\
	&=-\frac{d^2b^4}{S}\!\sum_{{{\bf r}_j},{{\bf r}'_j}}\!\frac{\exp[i({\bf q}_1\!-\!{\bf q}'_1){\bf r}_j\!+\!i({\bf q}_2\!-\!{\bf q}'_2){\bf r}'_{j}]}{|{\bf r}_j-{\bf r}'_{j}|^3},
\end{split}
\end{equation}
where $V({\bf r}_1-{\bf r}_2)$ is given by Eq. (\ref{eq:V}) of the main text, and $S$ is the surface area.
The last line of Eq. (\ref{Bf}) is obtained assuming the tight-binding regime, where the single particle wavefunction is
\begin{equation}\label{psis}
	\chi_{{\bf q}}({\bf r})=\frac{1}{\sqrt{N}}\sum_j\Phi_0({\bf r}-{\bf r}_j)\exp\left[i{\bf q}{\bf r}_j\right].
\end{equation}
Here, the index $j$ labels the lattice sites located at the points ${\bf r}_j$, and $N=S/b^2$ is the total number of the sites.
The particle wavefunction in a given site $j$ has extension $\xi_0$ and is expressed as $\Phi_0({\bf r}-{\bf r}_j)=(1/\sqrt{\pi}\xi_0)\exp[-({\bf r}-{\bf r}_j)^2/2\xi_0^2]$. 
In the low-momentum limit we may replace the summation over $j$ and $j'$ by the integration over $d^2r_j$ and $d^2r'_{j}$ taking into account that $b^2\sum_j$ transforms into $\int d^2r_j$. 
This immediately yields
\begin{equation}\label{flm}
	f({\bf k}',{\bf k})=-d^2\int\exp[i({\bf k}-{\bf k}'){\bf r}]\frac{d^2r}{r^3},
\end{equation}
and the $p$-wave part of the scattering amplitude is obtained multiplying Eq. (\ref{flm}) by $\exp(-i\phi)$ and integrating over $d\phi/2\pi$, 
where $\phi$ is the angle between the vectors ${\bf k}$ and ${\bf k}'$. 
This is the same result as in free space (see, e.g., \cite{Shlyapnikov}). 
The on-shell amplitude ($k=k'$) can be written as $f(k)=-(8\hbar^2/3m^*)kr^*_{\rm eff}$, 
where $r^*_{\rm eff}=m^*d^2/\hbar^2$ is the effective dipole-dipole distance in the lattice. 
The applicability of the Born approximation assumes that $kr^*_{\rm eff}\ll{1}$, 
which is clearly seen by calculating the second order correction to the scattering amplitude.

Up to the terms $\sim(kr^*_{\rm eff})^2$, 
the on-shell scattering amplitude following form the solution of the scattering problem for particles with mass $m^{*}$, is given by \cite{Shlyapnikov}:
\begin{equation}\label{eq:scattS}
	f(k)=-\frac{8}{3}\frac{\hbar^2}{m}kr^{*}+\frac{\pi}{2}\frac{\hbar^2}{m}(kr^{*})^2\ln\left({Bkr^{*}}\right),
\end{equation}
where the numerical coefficient $B$ comes form short-range physics. 
For calculating $B$ we introduce a perfectly reflecting wall at intermolecular distances ${r}{\sim}{b}$,
which takes into account that two fermions practically can not get to one and the same lattice site. 
The coefficient $B$ depends on the ratio $r^*_{\rm eff}/b$, and we show this dependence in Fig. 2a.

The treatment of the superfluid pairing is the same as in Ref. \cite{Shlyapnikov}, 
including the Gorkov-Melik-Barkhudarov correction.
We should only replace the mass $m$ with $m^{*}$. 
The expression for the critical temperature then becomes:
\begin{equation}\label{Seq:TcP}
	T_c=E_F\frac{\kappa}{(k_Fr^{*}_{\rm eff})^{9\pi^2/64}}\exp\left[-\frac{3\pi}{4k_Fr^{*}_{\rm eff}}\right],
\end{equation}
where $\kappa\simeq0.19B^{-9\pi^2/64}$,
and it is displayed in Fig. \ref{fig2}b as a function of $r^*_{\rm eff}/b$.

\subsubsection{Superfluid pairing of \\ fermionic polar molecules in a bilayer system}

For the interlayer interaction potential $V_L(r)$ given by equation (\ref{eq:potential}) 
in the main text, the scattering amplitude for $k_Fr^{*}{\ll}1$ can be calculated in the Born approximation~\cite{Santos}.
The $p$-wave part of the first order contribution to the off-shell amplitude is
\begin{equation}\label{eq:Sscatamplb}
	\begin{split}
	f_{1}(k',k)&=\int{\frac{d\varphi_k'd\varphi_k}{(2\pi)^2}e^{i(\varphi_k'-\varphi_k)}V_L(r)e^{i({\bf k}'-{\bf k})r}} \\ 
	&=-\frac{2\pi\hbar^2}{m}(kr^{*})\mathcal{F}_{1}(k',k,L),
	\end{split}
\end{equation}
where
\begin{equation}\label{eq:Sscatamplb2}
	\begin{split}
		\mathcal{F}_{1}(k'L,kL)=\frac{1}{kL}\int_{0}^{\infty}&{xdx\,J_1(k'Lx)J_1(kLx)} \\
		&\times{\frac{x^2-2}{(x^2+1)^{{5}/{2}}}},
	\end{split}
\end{equation}
and $J_1$ is the Bessel function.
Regarding the second order contribution, for the solution of the gap equation we only need the on-shell $p$-wave part,
which is given by
\begin{equation}\label{eq:Sscatamplb2}
	f_{2}(k)=\frac{2\pi\hbar^2}{m}(kr^{*})^2\mathcal{F}_{2}(kL),
\end{equation}
where 
\begin{equation}\label{eq:Sscatamplb3}
\begin{split}
	\!\!\!\!\mathcal{F}_{2}&(kL)=\frac{\pi}{\left(kL\right)^2}{\int}_{0}^{\infty}xdx\,J_1^2(kLx)\frac{x^2-2}{(x^2+1)^{{5}/{2}}}\\
	&\times{\int}_{x}^{\infty}{ydy\,J_1(kLy)N_1(kLy)\frac{y^2-2}{(y^2+1)^{{5}/{2}}}}.
\end{split}
\end{equation}
In fact, the true $p$-wave scattering amplitude follows from the exact relation
\begin{equation}
	f(k',k)=\int_0^{\infty}{J_1(k'r)V_L(r)\psi(k,r)2\pi rdr},  
\end{equation}
where $\psi(k,r)$ is the true wavefunction of the $p$-wave relative motion with momentum $k$, 
normalized such that for $r\rightarrow\infty$ we have $\psi(k,r)=J_1(kr)-(imf(k)/4\hbar^2)H^{(1)}_{1}(kr)$ with $H^{(1)}_{1}$ being the Hankel function. 
This amplitude is complex and it is related to the real amplitude $\tilde f=f_1+f_2$ given by equations (\ref{eq:Sscatamplb})--(\ref{eq:Sscatamplb3}) as 
\begin{equation}\label{eq:Sscatamplb4}
	f(k',k)=\frac{\tilde f(k',k)}{1+im\tilde f(k)/4\hbar^2}.  
\end{equation}

In order to calculate the superfluid transition temperature we use the BCS approach along the lines of Ref.~\cite{Shlyapnikov}.
We consider temperatures $T$ tending to $T_c$ from below and rely on the renormalized gap equation (\ref{eq:gap}).
For the $p$-wave pairing the order parameter is $\Delta_{\bf k}=\Delta(k)e^{i\varphi_k}$, 
and we then multiply Eq.~(\ref{eq:gap}) by $e^{-i\varphi_k}$ and integrate over $d\varphi_{k'}$ and $d\varphi_{k}$.
As a result, we obtain the same equation (\ref{eq:gap}) in which $\Delta_{{\bf k}}$ and $\Delta_{{\bf k}'}$ are replaced with $\Delta(k)$ and $\Delta(k')$, 
the off-shell scattering amplitude $f({\bf k}',{\bf k})$ with its $p$-wave part, 
and $\delta V({\bf k}',{\bf k})$ with its $p$-wave part $\delta{V}(k',k){=}\int{\delta{V}({\bf k'},{\bf k})\exp\left[i(\varphi_{k'}-\varphi_k)\right]d\varphi_{k'}d\varphi_{k}/4\pi^2}$. 
Calculating the contribution of the pole in the second term in square brackets and using Eq. (\ref{eq:Sscatamplb4}) we obtain
 \begin{eqnarray}\label{eq:gap_b}
	&\Delta(k)={-P}{\int}\frac{mdE_{k'}}{2\pi\hbar^2}\tilde{f}(k',k)\Delta(k')\!\left[\mathcal{K}(k'){-}\frac{1}{2(E_{k'}{-}E_k)}\right] \nonumber \\
	&-\int\frac{mdE_{k'}}{2\pi\hbar^2}\delta{V}(k',k)\Delta(k')\mathcal{K}(k'),
\end{eqnarray}
where the symbol $P$ stands for the principal value of the integral.
In the first term in the right hand side of Eq.~(\ref{eq:gap_b}) we divide the region of integration into two parts: $|E_{k'}-E_F|<\omega$ and $|E_{k'}-E_F|{>}\omega$,
where $\Delta(k_F),T_c\ll\omega\ll{E_F}$. 
The contribution to the $p$-wave order parameter from the first region we denote as $\Delta^{(1)}(k)$, 
and the contribution from the second region as $\Delta^{(2)}(k)$. 
The contribution of the second term in right hand side of equation (\ref{eq:gap_b}) is denoted as $\Delta^{(3)}(k)$.

\begin{figure*}[t]
\includegraphics[width=2\columnwidth]{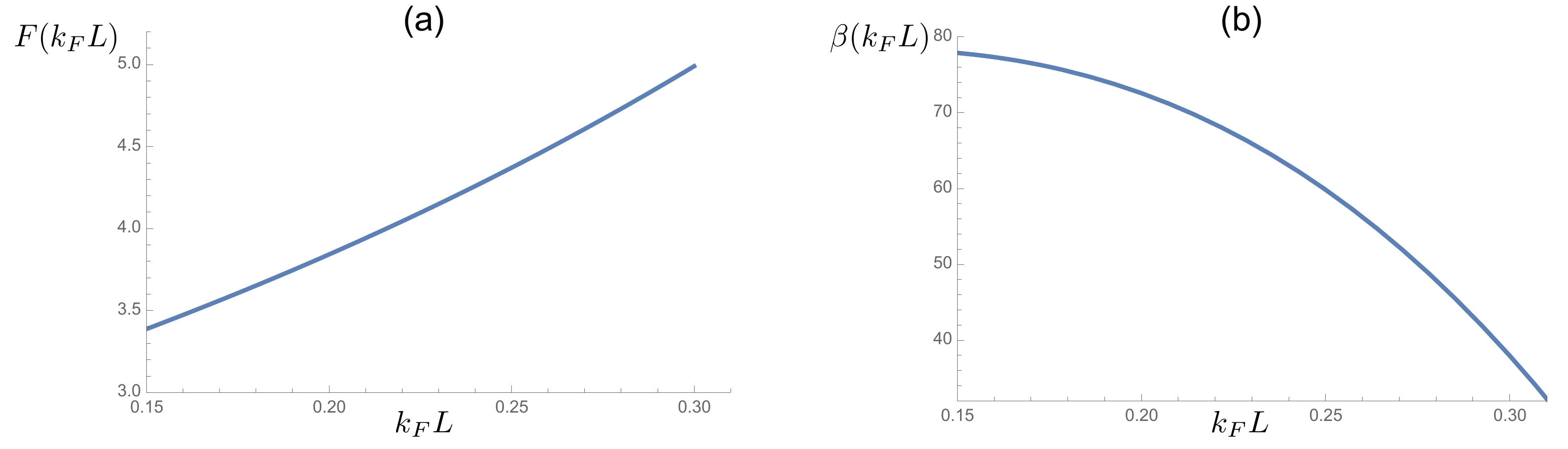}
\vskip -5mm
\caption
{
	The dependence of $F$ and $\beta$ on $k_FL$.
}
\label{fig3}
\end{figure*}

We first notice that the main contribution to $\Delta(k)$ comes from $k'$ close to $k_F$.
Retaining only $f_{1}$, which is proportional to $kr^{*}$, in the off-shell scattering amplitude and omitting the second term in the right hand side of Eq. (\ref{eq:gap_b})  
(which is proportional to $(kr^{*})^2$) we obtain
 \begin{equation}\label{eq:gap_b2}
	\Delta(k)=\Delta(k_F){f_{1}(k_F,k)}/{f_{1}(k_F)}.
\end{equation}
Putting $k=k_F$ and performing the integration in the first region in the first term of Eq~(\ref{eq:gap_b}), 
where ${E_F-\omega}<{E_{k'}}<{E_F+\omega}$,
we may put $\Delta(k'){=}\Delta(k_F)$ and $\tilde{f}(k',k)=\tilde{f}(k_F)=f_{1}(k_F)+f_{2}(k_F)$.
Then, putting $\epsilon_{k'}=|\xi_{k'}|$ in $\mathcal{K}(k')$ and taking into account that the contribution of the second term in square brackets is zero, 
we obtain:
\begin{equation}\label{eq:gap1}
	\Delta^{(1)}(k_F)=-\Delta(k_F)\rho(k_F)\tilde{f}(k_F)\ln\left(\frac{2e^{C}\omega}{\pi T_c}\right),
\end{equation}
with $C{=}0.577$ being the Euler constant, and $\rho(k_F)=m/2\pi\hbar^2$ the density of states. 

In the second region, where $E_{k'}{>}E_F{+}\omega$ or $E_{k'}<E_F-\omega$,
we put $\mathcal{K}=1/2|\xi_{k'}|$ and keep only $f_1$ in the scattering amplitude. 
For $k=k_F$ the integral over $E_{k'}$ from $E_F{+}\omega$ to $\infty$ vanishes. 
In the integral from $0$ to $E_F-\omega$ we use $\Delta(k')$ from Eq. (\ref{eq:gap_b2}) and find
\begin{equation}\label{eq:gap2}
\begin{split}
	\Delta^{(2)}(k_F)=-&\Delta(k_F)\rho(k_F)f_{1}(k_F) \\
	&\times\left[\ln\left(\frac{E_F}{\omega}\right)-\eta(k_FL)\right],
\end{split}
\end{equation}
where 
\begin{equation}
	\begin{split}
		\eta&(k_FL)=-2\int_0^{E_F-\omega}{\frac{k'dk'}{(k_F^2-k'^2)}\left[\frac{f_{1}^{2}(k_F,k)}{f_{1}^{2}(k_F)}-1\right]} \\	
		&=-2\int_0^{1}{\frac{ydy}{1-y^2}\left\{\left[\frac{\mathcal{F}_1(k_FL, k_FLy)}{\mathcal{F}_1(k_FL)}\right]^{2}-1\right\}},
	\end{split}
\end{equation}
and $\mathcal{F}_{1}(k_FL)\equiv\mathcal{F}_{1}(k_FL,k_FL)$

Then, we consider the Gor'kov-Melik-Barkhudarov corrections to the bare interaction of the molecules in the bilayer.
These many-body corrections are second order in $(k_Fr^{*})$ and are described by four diagrams (for details, see~\cite{GMB,Shlyapnikov,Baranov}). 
For the case of $p$-wave superfluidity of identical fermionic polar molecules they have been considered in Ref.~\cite{Shlyapnikov}.
They have been also studied for the interlayer $s$-wave superfluidity of dipoles oriented in the same direction in Ref.~\cite{Baranov}.

We are interested in the case of sufficiently small $k_FL$. 
Following the same treatment as in Refs. \cite{Shlyapnikov, Baranov}, in the limit of $k_FL\to0$ we obtain:
\begin{equation}
	\delta{V(k_F,k_F)}=-{\alpha\frac{\hbar^2}{m}(k_Fr^{*})^2},
\end{equation}
where $\alpha{\simeq}{10.57}$.
The dominant contribution to this result comes from the diagram containing a bubble in the interaction line (diagram a) in Refs. \cite{Shlyapnikov,Baranov}).
This contribution strongly decreases with increasing $k_FL$. 
In particular, for $k_FL\simeq0.15$ we have $\alpha\simeq{2.8}$, and $\alpha\simeq{2.2}$ when increasing $k_FL$ to $0.2$. 
Comparing $\delta{V}$ with the scattering amplitude $f_{1}(k_F)$ we see that for not very small $k_Fr^{*}$ the perturbative treatment of the Gor'kov-Melik-Barkhudarov corrections is 
adequate for $k_FL\gtrsim{0.15}$.
We therefore confine ourselves to these values of $k_FL$. 

Performing the integration in the second term of Eq.~(\ref{eq:gap_b}) we obtain the contribution of the Gor'kov-Melik-Barkhudarov corrections to the order parameter:
\begin{equation}\label{eq:gap3}
	\Delta^{(3)}(k_F)=\Delta(k_F)\frac{\alpha(k_FL)}{2\pi}(k_Fr^{*})^2\ln\left(\frac{E_F}{T_c}\right),
\end{equation}

The sum of Eqs. (\ref{eq:gap1}), (\ref{eq:gap2}) and (\ref{eq:gap3}) yields 
\begin{eqnarray}\label{eq:gap_c}
	&\Delta(k_F)=\Delta(k_F)\bigg[\bigg\{(k_Fr^{*})\mathcal{F}_{1}(k_FL)\ln\left(\frac{2e^{C-\eta(k_FL)}}{\pi}\frac{E_F}{T_c}\right)\bigg\} \nonumber \\
	&-\left\{\left(\mathcal{F}_{2}(k_FL){-}\frac{\alpha(k_FL)}{2\pi}\right)(k_Fr^{*})^2\ln\left(\frac{E_F}{T_c}\right)\right\}\bigg],
\end{eqnarray}
where we put $\omega\sim{E_F}$ in the terms proportional to $(k_Fr^{*})^2$.
We should also recall that the bare mass $m$ should be replaced with the effective mass $m^{*}=m[1-$ $(4/3\pi)k_Fr^{*}]$ which has been found in Refs.~\cite{Baranov,Lu}.
Since the relative difference between $m^{*}$ and $m$ is small as $k_Fr^{*}$, it is sufficient to replace $m$ with $m^{*}$ only in the multiple $r^{*}{\sim}{m}$ in the first term of Eq.~(\ref{eq:gap_c}).
This leads to the appearance of a new term
\begin{equation}
	-\Delta(k_F)\frac{4}{3\pi}\mathcal{F}_{1}(k_FL)(k_Fr^{*})^2\ln\left(\frac{E_F}{T_c}\right)
\end{equation}
in the right hand side of equation (\ref{eq:gap_c}).
Then, dividing both sides of Eq.~(\ref{eq:gap_c}) by $\Delta(k_F)$ we obtain for the critical temperature:
\begin{equation}\label{eq:Tcbilayer}
	T_c=E_F\beta(k_FL)\exp\left[-\frac{F(k_FL)}{k_Fr^{*}}\right],
\end{equation}
where 
\begin{equation}
	F(k_FL)=\left[{\mathcal{F}_{1}(k_FL)}\right]^{-1}
\end{equation}
and 
\begin{equation}
	\begin{split}
	&\!\!\beta(k_FL)=\exp\bigg[C+\ln\left(\frac{2}{\pi}\right)-\eta(k_FL)\\ 
	&-\frac{\mathcal{F}_{2}(k_FL)}{\mathcal{F}_{1}^{2}(k_FL)}-\frac{4}{3\pi}\frac{1}{\mathcal{F}_{1}(k_FL)}+\frac{\alpha(k_FL)}{2\pi}F^2(k_FL)\bigg].
	\end{split}
\end{equation}
The dependence of $F$ and $\beta$ on $k_FL$ is shown in Fig.~3.
We stop at $k_FL=0.3$ because for larger values of this parameter the function $F$ is so large that the critical temperature will be negligible. 

\subsection{Acknowledgments}
We acknowledge fruitful discussions with Jun Ye, Ignacio Cirac, and Martin Zwierlein.
The research leading to these results has received funding from the 
European Research Council under European Community's Seventh Framework Programme (FR7/2007-2013 Grant Agreement no. 341197).

\medskip
\noindent
{\bf Author contributions}:~A.K. Fedorov, S.I. Matveenko, and V.I. Yudson equally contributed to the obtained results. 
G.V. Shlyapnikov supervised the project, and A.K. Fedorov and G.V. Shlyapnikov have written the text.

\noindent
{\bf  Competing financial interests}:~The authors declare no competing financial interests.

\noindent
{\bf  Corresponding author}:~Correspondence should be addressed to A.K. Fedorov (akf@rqc.ru).

\end{document}